\documentclass{aastex631}
\usepackage {subfigure}
\usepackage{rotating}
\usepackage{longtable}
\usepackage {multirow}
\usepackage{pifont}

\shortauthors{Deng et al.}
\graphicspath{{./}{figures/}}

\begin{document}

\title{On the magnetic braking law in black hole low-mass X-ray binaries}

\author[0000-0002-1398-2694]{Zhu-Ling Deng}
\affil{School of Astronomy and Space Science, Nanjing University, Nanjing 210023, China; lixd@nju.edu.cn }
\affil{Key Laboratory of Modern Astronomy and Astrophysics (Nanjing University), Ministry of Education, Nanjing 210023, China}
\author[0000-0002-0584-8145]{Xiang-Dong Li}
\affil{School of Astronomy and Space Science, Nanjing University, Nanjing 210023, China; lixd@nju.edu.cn }
\affil{Key Laboratory of Modern Astronomy and Astrophysics (Nanjing University), Ministry of Education, Nanjing 210023, China}

\begin{abstract}
Magnetic braking (MB) plays an important role in the evolution of close low-mass X-ray binaries (LMXBs). It is also essential to the formation of ultracompact X-ray binaries (UCXBs). There have been lively investigations on the MB mechanism(s) in both single stars and close binaries including cataclysmic variables and neutron star (NS) LMXBs, but with diverse conclusions. In this paper we explore the effect of MB on the black hole (BH) LMXB evolution. We combine binary population synthesis with detailed binary evolution to obtain
the expected properties of Galactic BH LMXB population. The simulated results are compared with the observational data including the BH mass, companion mass, companion temperature, orbital period, and mean accretion rate. Our results reveal that the MB laws with relatively low efficiency (i.e., RM12 and RVJ83) exhibit better agreement with observations, contrary to what was found for NS LMXBs. This raises the interesting question about whether MB really follows the same unified law in different types of binaries. We also predict that only a very small fraction ($\lesssim 2.5\%$) of BH LMXBs can evolve to be UCXBs. This explains why there is no BH UCXB discovered by far.

\end{abstract}


\section{Introduction} 
There are 23 black hole intermediate/low-mass X-ray binaries (BH I/LMXBs) discovered in the Milky Way \citep[][]{McClintock2006,Belloni2011,Corral2016}. Table 1 lists their main parameters.  The orbital periods range from  around 0.1 day (0.107 day for Swift J1375$-$0933, \citealt{Mata2015}, \citealt{Casares2022}; 0.1 day for MAXI J1659$-$152, \citealt{Yamaoka2012}, \citealt{Kuulkers2013}, \citealt{Torres2021}; and 0.092 day for MAXI J0637$-$430, \citealt{Soria2022}) to around 34 days \citep[GRS 1915$+$105,][]{Greiner2001Nat}. About two-thirds of them are close binaries with orbital periods less than 1 day. In these binaries the companions are mostly K or M-type main-sequence (MS) stars, while in binaries with orbital periods longer than 1 day, the companions are mostly giant stars \citep{Li2015}. Almost all BH LMXBs are transient sources with quiescent X-ray luminosities $\sim 10^{30}-10^{34}$ erg\,s$^{-1}$ \citep{Garcia2001,Carotenuto2022}, but the luminosities can increase by several orders of magnitude during outbursts, reaching $\sim 10^{37}-10^{39}$ erg\,s$^{-1}$. The BH masses measured by dynamical methods are distributed between $\sim 3\,M_{\odot}$ and $>15\,M_{\odot}$ \citep{Casares2014}. Studies of the mass distribution of BHs and neutron stars (NSs) show that there is a gap around $2.5-5\,M_{\odot}$ between these two types of objects \citep{Bailyn1998,Ozel2010,Farr2011}. Currently, the spins of $\sim$ 20 BHs have been measured through the X-ray continuum-fitting  \citep{Zhang1997} and relativistic reflection \citep{Fabian1989} methods. 
They are distributed from $0.12\pm0.19$ \citep[A0620$-$00,][]{Gou2010} to $>0.99$ \citep[4U 1957$+$115,][]{Barillier2023,Marra2023}.

In the standard model, BH LMXBs evolve from primordial binaries with extreme mass ratios. When the more massive primary star evolves to be a supergiant and overflows its Roche lobe (RL), the mass transfer is usually dynamically unstable, leading to common envelope evolution \citep{Paczynski1976}. If the secondary star can successfully eject the envelope of the primary star, a short orbital period binary consisting of the core of the primary star and the secondary star is left. The core then collapses and gives birth to a BH. If the supernova (SN) explosion during this process does not disrupt the binary system, it will evolve to be a BH intermediate/low-mass X-ray binary (I/LMXB). The main difficulty for this evolutionary channel is that the primordial binary is in short of enough orbital energy to eject the envelope of the primary star \citep{Portegies1997,Kalogera1999,Podsiadlowski2003}, causing merger of the two stars. Therefore, there are alternative scenarios proposed for the formation of BH LMXBs in the literature. For example,  there could be additional energy to help eject the primary’s envelope \citep{Ivanova2002,Podsiadlowski2010}, or BH LMXBs have actually evolved from IMXBs with extra angular momentum loss mechanisms \citep{Justham2006,ChenWC2006}, from triple star systems \citep{Eggleton1986}, from Thorne-$\dot{\rm Z}$ytkow objects \citep{Podsiadlowski1995}, and from binaries experienced failed SN explosions \citep{Wang2016MN,Shao2019,Shao2020}. Thus, there are still many uncertainties regarding the formation of BH LMXBs.

After the formation of a LMXB, the orbital evolution and mass transfer are driven by orbital orbital angular momentum loss if its orbital period is shorter than a few days \citep{Ma2009,Istrate2014a,Pavlovskii2016,Van2019a,Van2019b,Deng2021}. In addition to mass loss, gravitational radiation (GR) and magnetic braking (MB) dominate angular momentum loss in binary systems with orbital periods shorter and longer than a few hours, respectively. The idea of MB originates from the observations of the spin deceleration of solar-like stars \citep{Skumanich1972},  which is attributed to the coupling of the stellar wind and magnetic field that takes away the stellar spin angular momentum \citep{Mestel1968}. If the binary orbital revolution and the spin of the secondary star are synchronized, MB can extract the orbital angular momentum of the binary system \citep{King1988}.

\citet{Verbunt1981} and \citet{Rappaport1983} derived empirical formula of MB based on the \cite{Skumanich1972} model, and applied it to LMXBs, which is currently the most widely used MB prescription (RVJ83). However, it faces severe difficulties in explaining the following facts: (1) The observed accretion rates of some NS LMXBs is about $1-2$ orders of magnitude higher than predicted  \citep{Podsiadlowski2002,Pfahl2003,Shao2015, Pavlovskii2016,Van2019a,Deng2021}. (2) It is difficult to form Ultracompact X-ray binaries (UCXBs) within the Hubble time  \citep{Podsiadlowski2002,vanderSluys2005}. (3) The predicted orbital distribution of binary pulsars (which are the evolutionary products of NS LMXBs) is not compatible with the observations \citep{Pfahl2003,Istrate2014a,Shao2015}. These stimulated modified versions of the MB law for LMXBs \citep{Pavlovskii2016,Van2019a,Van2019b}. In the mean time, based on the study of the spin evolution of low-mass stars (either single or in binaries), various types of MB laws have also been proposed \citep[e.g.,][]{Sills2000,Matt2012,Reiners2012,Gallet2013,vanSaders2013,SadeghiArdestani2017}.

To examine and discriminate the efficiency of different MB laws, \citet{Deng2021} investigated the LMXB evolution with five representative MB laws, i.e., the RVJ83 law, the $\tau$-boosted law \citep[which is the RVJ83 law with the addition of a convection boosted facter,][]{Van2019a}, the convection and rotation boosted law \citep[or VI19 law;][]{Van2019b}, the Matt12 law \citep[based on 2D axisymmetric magnetohydrodynamical simulation of wind outflows from a rigidly rotating star with an aligned dipolar field;][]{Matt2008,Matt2012}, and the RM12 law \citep[which assumes the stellar magnetic field reaches saturation when the surface angular velocity is above a critical value;][]{Reiners2012}. Comparison between theory with observation seems to favor the $\tau$-boosted and VI19 laws. It is then interesting to see whether the same conclusion holds for BH LMXBs. 

In this work, we combine binary population synthesis (BPS) with detailed stellar evolution calculation to study the formation and evolution of BH LMXBs with different MB laws. This paper is organized as follows. Section 2 describes the binary evolution models and compares the LMXB evolution with different MB laws. Section 3 presents the population properties by combining the BPS model with the binary evolution models. We examine the influence of the SN mechanisms and the MB laws on the formation of LMXBs and UCXBs. Our concluding remarks are in Section 4.

\begin{deluxetable*}{cccccccc}
\tablenum{1}
\tablecaption{The data of observed BH L/IMXBs in the Milky Way}
\tablehead{
Source &  $M_{\rm BH}$ (M$_{\odot}$) & $M_{\rm c}$ (M$_{\odot}$) & $P_{\rm orb}$ (days) & Spectral Type  & $T_{\rm eff}$ & $a_{\star}$ & $\dot{M}_{\rm tr}$ (M$_{\odot}$\,yr$^{-1}$)}
\startdata
	GRS 1915+105 (1) & 12.4$\pm$2.0 & 0.58$\pm$0.33 & 33.85 & K0III-K3III & 4398-4612 & 0.95$\pm$0.05 & $<1.5\times 10^{-7}$  \\
	GS 2023+338 (2) & 9.0$\pm$0.6 & 0.54$\pm$0.05 & 6.47 & K0III-K3III & 4398-4612 & ... & $9.3\times 10^{-10}$  \\
	V4641 Sgr (3) & 6.4$\pm$0.6 & 2.9$\pm$0.4 & 2.82 & B9III & 12261-12461 & ... & ...  \\
	GRO J1655-40 (4) & 5.31$\pm$0.07 & 1.75$\pm$0.25 & 2.62 & F5III-F7III & 5715-5990 & 0.7$\pm$0.1 & $8.7\times 10^{-10}$ \\
	GS 1354-64 (5) & 7.6$\pm$0.7 & $\sim$1.03 & 2.54 & G0III-G5III & 4963-5382 & ... & $> 1.6\times 10^{-9}$ \\
	GX 339-4 (6) & 9$^{+1.6}_{-1.2}$ & 0.7$\pm$0.4 & 1.75 & K1V & 4954-5154 & 0.95$^{+0.02}_{-0.08}$ & $1.8\times 10^{-8}$ \\
	XTE J1550-564 (7) & 9.1$\pm$0.6 & 0.3$\pm$0.07 & 1.54 & K3III/V & 4398-4786 & 0.49$^{+0.13}_{-0.2}$ & $1.6\times 10^{-9}$ \\
	4U 1543-47 (8) & 9.4$\pm$2.0 & 2.7$\pm$1.0 & 1.12 & A2V & 9000$\pm$500 & 0.8$\pm$0.1 & $8.7\times 10^{-10}$ \\
	MAXI J1820+070 (9) & 5.7-8.3 & 0.3-0.8 & 0.69 & K3V-K5V & 4500-4786 & 0.14$\pm$0.09 & ... \\
	H1705-250 (10) & 6.4$\pm$1.5 & 0.25$\pm$0.17 & 0.52 & K3V-M0V & 3739-4786 & ... & < $2.8\times 10^{-10}$ \\
	GRS 1124-68 (11) & 11.0$^{+2.1}_{-1.4}$ & 0.89$^{+0.18}_{-0.11}$ & 0.43 & K3V-K5V & 4500-4786 & 0.63$^{+0.16}_{-0.19}$ & <$3.3\times 10^{-10}$ \\
	MAXI J1305-704(12) & 8.9$^{+1.6}_{-1.0}$ & 0.43$\pm$0.16 & 0.394 & K3V-K5V & 4450-4740 & ... & ... \\
	4U 1957+115$^{BHC}$ (13) & 4.6$\pm$0.46 & <1 & 0.39 & ... & ... & 0.992$\pm$ 0.001 & 5.67$\times 10^{-10}$ \\
	GS 2000+251 (14) & $\sim$6.55 & 0.16-0.47 & 0.34 & K3V-K6V & 4352–4786 & ... & <$6.5\times 10^{-11}$ \\
	A0620-00 (15) & 6.61$\pm$0.25 & 0.4$\pm$0.05 & 0.32 & K5V-K7V & 4201–4500 & 0.12$\pm$0.19 & $5.4\times 10^{-11}$ \\
	XTE J1650-500(16) & 5.65$\pm1.65$ & 0.56$\pm$0.16 & 0.32 & ... & ... & 0.79$\pm0.01$ & <$1.09\times 10^{-10}$ \\
	GRS 1009-45 (17) & 8.5$\pm$1.0 & 0.54$\pm$0.1 & 0.29 & K7V-M0V & 3739-4201 & ... & $<2.7\times 10^{-10}$ \\
	XTE J1859+226 (18) & 7.85$\pm$0.46 & 0.55$\pm$0.16 & 0.276 & K5V-K7V & 4201–4500 & 0.149$\pm$0.005 & ... \\
	GRO J0422+32 (19) & 2.7$^{+0.7}_{-0.5}$ & 0.33$^{+0.28}_{-0.20}$ & 0.212 & M1V–M4V & 3122–3584 & ... & <$2.7\times 10^{-11}$ \\
	XTE J1118+480 (20) & 7.6$\pm$0.7 & 0.18$\pm$0.07 & 0.17 & K7V-M1V & 3584-4202 & ... & $2.1\times 10^{-10}$ \\
	Swift J1375-0933$^{BHC}$ (21) & 10.9$^{+1.7}_{-1.6}$ & 0.42$^{+0.04}_{-0.03}$ & 0.107 & M2V-M6V & 2822-3429 & ... & ... \\
	MAXI J1659-152$^{BHC}$ (22) & 3.3-7.5 & 0.06-0.22 & 0.1 & M2V–M5V & 2971–3429 & 0.21$^{+0.14}_{-0.2}$ & $1.4\times 10^{-10}$ \\
	MAXI J0637-430$^{BHC}$ (23) & 5.1$\pm$1.6 & 0.25$\pm$0.07 & 0.092 & ... & ... & ... & ...
\enddata
\tablecomments{
 References. (1) \citet{Greiner2001Nat,Greiner2001AA,Harlaftis2004,McClintock2006,Steeghs2013,Reid2014} (2) \citet{Casares1994,Khargharia2010,Hynes2009}. (3) \citet{Orosz2001,Sadakane2006, MacDonald2014}. (4) \citet{Shafee2006,Gonzalez2008,Motta2014a}. (5) \citet{Casares2009}. (6) \citet{Cowley1987,Hynes2003,Steiner2013,Parker2016}. (7) \citet{Orosz2011ApJ,Steiner2011}. (8) \citet{Orosz1998,Orosz2003,Shafee2006}. (9) \citet{Torres2019,Torres2020,Zhao2021}. (10) \citet{Remillard1996,Harlaftis1997}. (11) \citet{Shahbaz1997,Gelino2001,Wu2015,Wu2016,Chen2016} (12) \citet{Mata2021}. (13) \citet{Barillier2023,Marra2023} (14) \citet{Chevalier1990,Filippenko1995,Casares1995,Barret1996,Harlaftis1996,Ioannou2004,Steiner2013}. (15) \citet{Neilsen2008,Johannsen2009,Cantrell2010,Gou2010}. (16) \citet{Orosz2004,Miller2009}. (17) \citet{Filippenko1999,Macias2011}. (18) \citet{Zurita2002,Motta2022,Yanes-Rizo2022}. (19) \citet{Harlaftis1999,Reynolds2007, Casares2022}. (20) \citet{Calvelo2009,Gonzalez2012,Khargharia2013}. (21) \citet{Mata2015, Casares2022}. (22) \citet{Yamaoka2012,Kuulkers2013,Torres2021,Feng2022}. (23) \citet{Soria2022}.
}
\end{deluxetable*}

\section{Modeling the Evolution of BHI/LMXBs: Methods and Results} 

\subsection{Methods and Initial Parameters}
We simulate the evolution of BH LMXBs with Modules for Experiments in Stellar Astrophysics \citep[MESA,][]{Paxton2011,Paxton2013,Paxton2015,Paxton2018,Paxton2019}. The initial binary system consists of a BH and a MS companion with solar chemical compositions ($X=0.7$, $Z=0.02$). The convective mixing parameter is set to $\alpha=2$, and convective overshooting and semiconvection are not considered. We take into account three mechanisms of orbital angular momentum loss: MB, mass loss, and GR. We adopt the \citet{Ritter1988} scheme to calculate the mass transfer rate via Roche lobe overflow (RLO), and set the maximum accretion rate onto the BH to be the Eddington limited accretion rate. When the mass transfer rate is super-Eddington, we assume that the excess material leaves the binary system in the form of an isotropic wind from the BH. In this work, we adopt four MB laws from \citet{Deng2021}, namely the RVJ83, RM12, $\tau$-boosted, and VI19 laws \citep{Rappaport1983,Reiners2012,Van2019a,Van2019b}. We refer to \citet{Deng2021} for more detailed descriptions of these models,  and 
emphasize that these MB laws are not mutually exclusive, but based on different assumptions, and therefore are unlikely to perfectly describe nature.

\subsection{Example Evolution with Different MB Laws}

We first discuss the impact of different MB laws from a theoretical perspective. Figure 1 shows the evolution of a BH LMXB with $M_{\rm BH}=8\,M_{\odot}$, $M_{\rm donor}=1\,M_{\odot}$, and $P_{\rm orb}=1$ day with the four MB laws. The left panel demonstrates the orbital evolution. The binary orbit in all the four cases shrinks due to MB, which causes the companion star to fill its RL and initiate mass transfer when the orbital period decreases to about 8 hours. The orbital evolution is determined by the competition between orbital shrinking mainly caused by angular momentum loss (due to MB and/or GR) and orbital expansion due to mass transfer. The VI19 MB law predicts significantly higher mass transfer rate (see the right panel) and hence longer orbital period around $0.2-0.3\,M_{\odot}$ than other laws. When the companion star becomes fully convective, MB ceases working and the subsequent orbital shrinking is dominated by GR. 

The middle panel compares the ratios of the MB-induced angular momentum loss rate and the total angular momentum loss rate with the four MB laws. Generally MB dominates angular momentum loss ($\dot{J}_{\rm mb}/\dot{J}_{\rm total}>0.5$) before the mass transfer. During the mass transfer, the MB efficiency with the RM12 law decreases all the way with decreasing orbital period, while the MB efficiencies with other laws first increase and then decrease/terminate until the donor star becomes fully convective. 

The right panel depicts the evolution of the mass transfer rate. With the VI19 law, the mass transfer proceeds most rapidly with the peak value of the mass transfer rate around $\sim 10^{-8}M_{\odot}$yr$^{-1}$, and it becomes lower in the order of the $\tau$-boosted, RVJ83, and RM12 laws. When the mass transfer rate is lower than a critical value, the accretion disk is regarded to be thermally and viscously unstable \citep{Lasota2001,Lasota2008}. The grey shaded line shows the critical mass transfer rate for disk instability with X-ray irradiation considered \citep{Lasota2008}. Before MB ceases, the accretion rates are higher than the critical value with the VI19 and $\tau$-boosted laws in most cases, but lower than the critical value with the RM12 law. The accretion rate with the RVJ83 law is close to the critical value. Thus, the occurrence of transient behaviour in BH LMXBs could be potentially used to discriminate the MB laws.

\begin{figure}
    \centering
    \includegraphics[width=16cm]{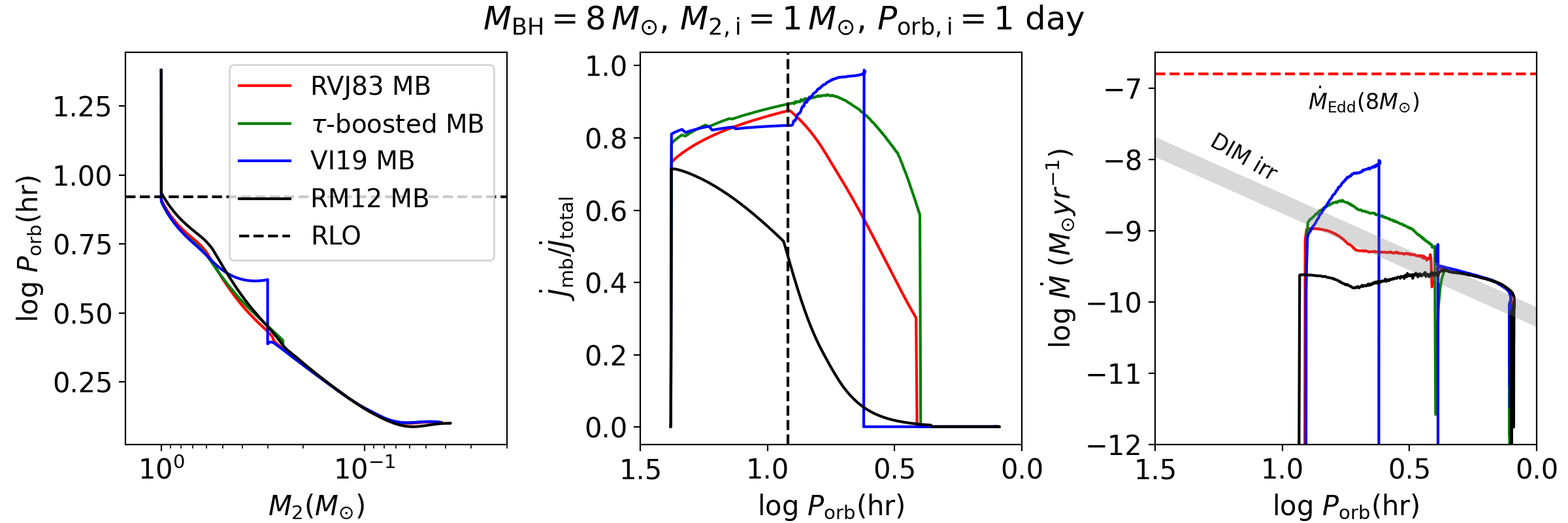}
\caption{Evolution of a BH LMXB by using different MB formulae. The initial BH mass, companion star mass, and orbital period of the binary system are 8$M_{\odot}$, 1.0$M_{\odot}$, and 1 day, respectively. The left, middle, and right panels show the evolution of the orbital period with companion star mass, the ratio of the magnetic braking-induced orbital angular momentum loss rate to the total orbital angular momentum loss rate as a function of the orbital period, and the accretion rate as a function of the orbital period, respectively. The red, green, blue, and black lines represent the RVJ83, $\tau$-boosted, VI19, and RM12 MB models, respectively. The black dashed line, red dashed line, and gray thick line denote the orbital period when the companion star fills its RL, the Eddington accretion rate of a $8\,M_{\odot}$ BH, and the critical mass transfer rate for disk instability under X-ray irradiation, respectively.
}
\end{figure}

\subsection{Population Properties and Comparison with Observations}

We then follow the  evolution of a large number of BH I/LMXBs. We set the initial BH mass to be $8\,M_{\odot}$, and the initial donor mass $0.8-7\,M_{\odot}$ in step of 0.2 $M_{\odot}$.  The logarithm of the initial binary orbital period is distributed in the range $-0.5\leq \log P_{\rm orb,i}({\rm day})\leq 2$ in step of $\Delta\log P_{\rm orb,i}({\rm day}) =0.05$. 

Figure 2 demonstrates the calculated binary evolutionary tracks with the initial donor mass of $1.0-3.0\,M_{\odot}$. Also shown are the measured values of the mean accretion rate, donor mass, spectral type, and orbital period of 15 BH LMXBs in Table 1. 
Note that the measured the mass and spectral type of the donor star through dynamical method and spectral fitting respectively are not always consistent with each other. The magnitudes of the calculated mass transfer rates are indicated by the color bar on the right, and the data of the observed mean accretion rates are taken from \citet{Coriat2012}. 

We can see that all the four MB laws are able to reproduce the companion star masses and orbital periods of the observed BH LMXBs, as well as the accretion rates of binaries with $P_{\rm orb}>1$ day. For binaries with $P_{\rm orb}<1$ day, the accretion rate calculated with the VI19, $\tau$-boosted and RVJ83 laws are a few times higher than observed, while the accretion rates with the RM12 law are basically consistent with observations.
Figure 2 also demonstrates that binaries with initial companion mass $\leq 1.4\,M_{\odot}$ will evolve to either narrower or wider systems, while binaries with initial companion mass $\geq 2.0\,M_{\odot}$ will expand all the way, independent of the MB laws.  

\begin{figure}
    \centering
    \includegraphics[width=18cm]{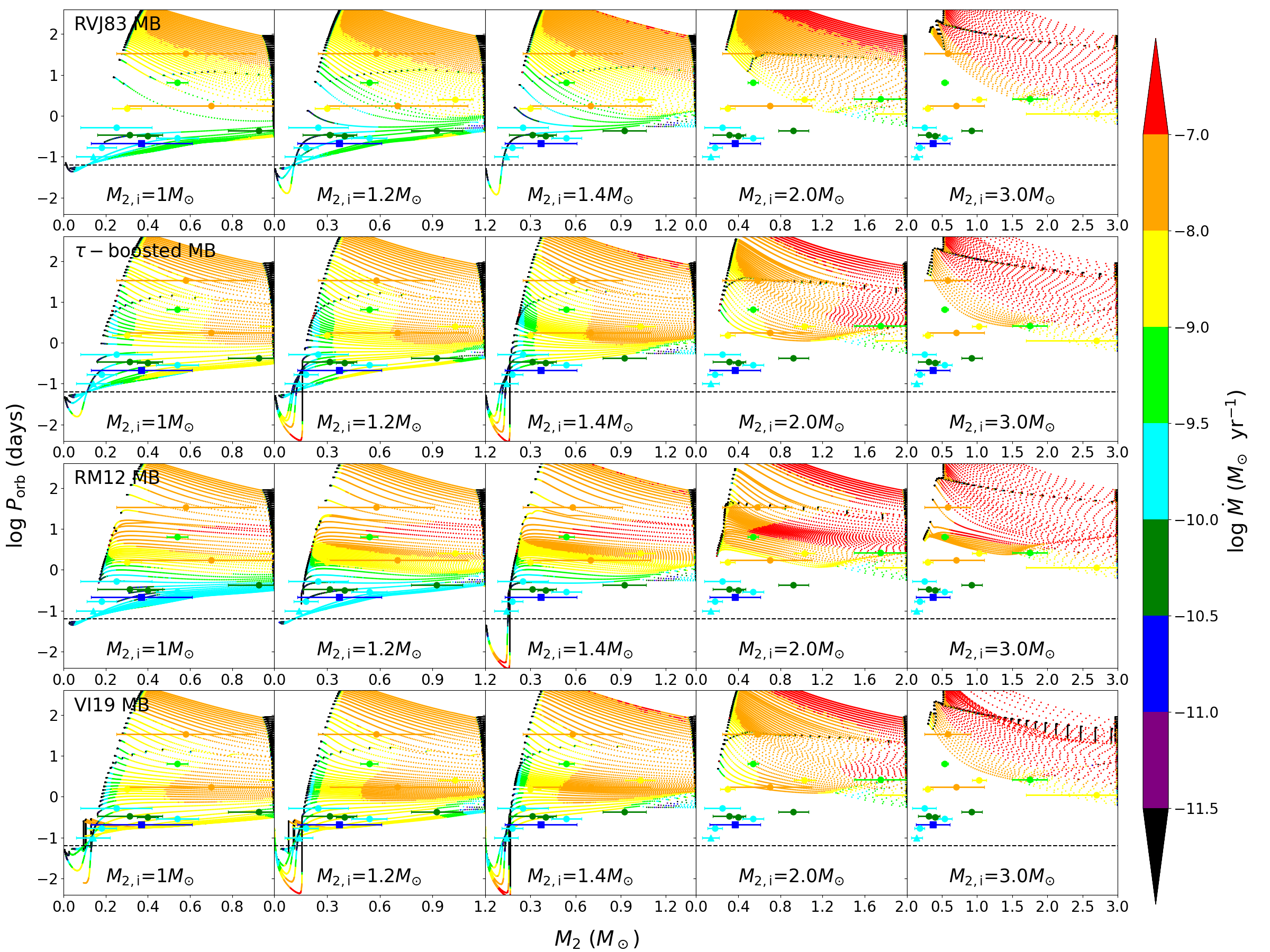}
    \caption{The evolution of BH I/LMXBs with different MB laws in the companion mass \textit{vs.} orbital period plane. The initial BH mass is $8\,M_{\odot}$. The let to right columns correspond to the initial companion masses of 1.0, 1.2, 1.4, 2.0, and $3.0\,M_{\odot}$, respectively. The top to bottom rows correspond to the RVJ83, $\tau$-boosted, RM12, and VI19 MB laws, respectively. Different colors represent the range of the mass transfer rate. The dots with error bars display the observed BH I/LMXBs. The square and triangle error bars represent the data for the sources GRO J0422+320422 and MAXI J1659-152, respectively. The black dashed line indicates the orbital period of 90 minutes. 
}

\end{figure}

Figure 3 shows the evolutionary tracks in the donor's effective temperature - orbital period plane. 
By comparing with the observations, one can see that the RVJ83, $\tau$-boosted, and RM12 laws match most BH LMXBs except GRO J0422+32 and MAXI J1659-152 \footnote{Here we do not consider the possible influence of X-ray irradiation on the companion star and optical contamination by the accretion disk}. The VI19 law seems able to account for the above two sources, but the predicted accretion rates do not align with the observations. 

\begin{figure}
    \centering
    \includegraphics[width=18cm]{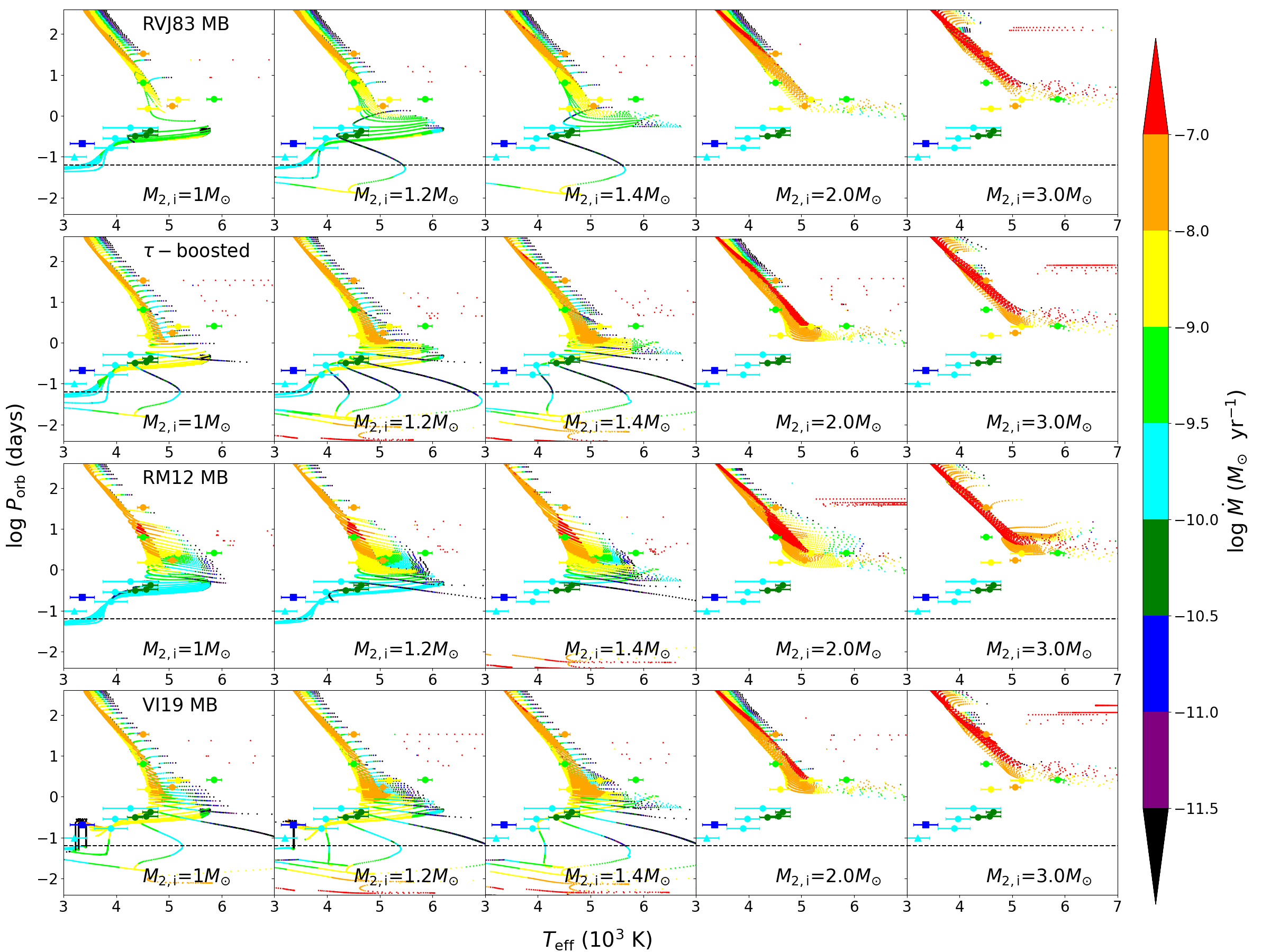}
    \caption{Similar to Figure 2, but for the relationship between the donor's temperature and the orbital period.}

\end{figure}

Table 2 quantitatively compare the effectiveness of the four MB laws in matching the observations of BH LMXBs. Since the donor's mass and spectral type for some systems in Table 1 are not exactly in accord with each other, we select two sets of parameters for comparison. The first set, denoted as $C_1$, include the BH mass, companion's temperature, orbital period, and accretion rate, and the second set $C_2$ add the companion mass. For the BH mass, donor mass, effective temperature, and orbital period, we consider successful match if the calculated values fall within $0.95-1.05$ of the observed data; for the accretion rate, the match means the calculated value falls within 0.5 dex of the logarithmic value of the inferred mean accretion rate. We use the symbols \ding{55}, \ding{52}, \ding{52}\ding{52}, and \ding{52}\ding{52}\ding{52} to denote that the BHI LMXB can be reproduced with 0, $1-4$, $5-9$, and $>9$ evolutionary tracks with a specific MB law, respectively.

According to Table 2, For set $C_1$, the RVJ83, $\tau$-boosted, and RM12 laws can reproduce 13 of the 15 selected BH I/LMXBs, while the VI19 law can reproduce only 8 systems. For set $C_2$, the RM12, $\tau$-boosted, RVJ83, and VI19 laws can match 11, 10, 9, and 7 BH I/LMXBs, respectively. In short, among the four MB laws used in this work, the RM12 law seem to perform best in accounting for the evolution of BH LMXBs. In addition, we found that GRO J0422$+$32 and MAXI J1659$-$152 cannot be reproduced with any MB law.

\begin{deluxetable*}{ccccccccc}
\tablenum{2}
\tablecaption{Comparison of the Efficiency of different MB laws in the evolution of BH I/LMXBs}
\tabletypesize{\scriptsize}
\tablehead{
 Source & \multicolumn{2}{c}{RVJ83}  & \multicolumn{2}{c}{$\tau$-boosted}  & \multicolumn{2}{c}{RM12}  & \multicolumn{2}{c}{VI19}  }
\startdata
 &C$_1$ & C$_2$ & C$_1$ & C$_2$ & C$_1$ & C$_2$ & C$_1$ & C$_2$ \\
\hline
GRS 1915+105 & \ding{52}\ding{52}\ding{52} & \ding{52}\ding{52}\ding{52} & \ding{52}\ding{52}\ding{52} & \ding{52}\ding{52}\ding{52} & \ding{52}\ding{52}\ding{52} & \ding{52}\ding{52}\ding{52} & \ding{52}\ding{52}\ding{52} & \ding{52}\ding{52}\ding{52}   \\
GS 2023+338 & \ding{52}\ding{52}\ding{52} & \ding{52}\ding{52}\ding{52} & \ding{52}\ding{52}\ding{52} & \ding{52}& \ding{52} & \ding{55} & \ding{52}\ding{52}\ding{52} & \ding{55} \\
GRO J1655-40 & \ding{52}\ding{52}\ding{52} & \ding{52}\ding{52}\ding{52} & \ding{52}\ding{52}\ding{52} & \ding{52}\ding{52}\ding{52} & \ding{52}\ding{52}\ding{52} & \ding{52}\ding{52} & \ding{52}\ding{52}\ding{52} & \ding{52}\ding{52}\ding{52}    \\
GS 1354-64 & \ding{52}\ding{52}\ding{52} & \ding{52}\ding{52}\ding{52} & \ding{52}\ding{52}\ding{52} & \ding{52} & \ding{52}\ding{52}\ding{52} & \ding{52} & \ding{52}\ding{52}\ding{52} &\ding{52}    \\
GX 339-4 & \ding{52}\ding{52}\ding{52} & \ding{55}\ & \ding{52}\ding{52}\ding{52} & \ding{52}\ding{52}\ding{52} & \ding{52}\ding{52}\ding{52} & \ding{52}\ding{52}\ding{52} & \ding{52}\ding{52}\ding{52} & \ding{52}\ding{52}\ding{52}    \\
XTE J1550-564 & \ding{52}\ding{52}\ding{52} & \ding{55} & \ding{52}\ding{52}\ding{52} & \ding{52}\ding{52}\ding{52} & \ding{52}\ding{52}\ding{52} & \ding{52}\ding{52}\ding{52} & \ding{52}\ding{52}\ding{52} & \ding{52}\ding{52}\ding{52}\\
4U 1543-47  & \ding{52}\ding{52}\ding{52}  & \ding{52}\ding{52}\ding{52} & \ding{52}\ding{52}\ding{52} & \ding{52}\ding{52}\ding{52} & \ding{52}\ding{52}\ding{52}  & \ding{52}\ding{52}\ding{52} & \ding{52}\ding{52}\ding{52} & \ding{52}\ding{52}\ding{52}   \\
H1705-250 & \ding{52} & \ding{55} & \ding{52}\ding{52} & \ding{52}\ding{52} & \ding{52}\ding{52}\ding{52} & \ding{52}\ding{52}\ding{52} & \ding{52} & \ding{52}    \\
GRS 1124-68  & \ding{52} & \ding{55} & \ding{52} & \ding{55} & \ding{52}\ding{52}  & \ding{55} & \ding{55}  & \ding{55}   \\
GS 2000+251 & \ding{52}\ding{52} & \ding{52}\ding{52} & \ding{52}& \ding{52} & \ding{52}\ding{52}& \ding{52}\ding{52} & \ding{55} & \ding{55}  \\
A0620-00 & \ding{52} & \ding{52} & \ding{52} & \ding{55} & \ding{52}\ding{52}& \ding{52} & \ding{55} & \ding{55}  \\
GRS 1009-45 & \ding{52}\ding{52}\ding{52}& \ding{52}\ding{52} & \ding{52}\ding{52}\ding{52} & \ding{55} & \ding{52}\ding{52}\ding{52}& \ding{52}\ding{52} & \ding{55} & \ding{55}   \\
GRO J0422+32 & \ding{55} & \ding{55} & \ding{55} & \ding{55} & \ding{55} & \ding{55} & \ding{55} & \ding{55}   \\
XTE J1118+480 & \ding{52}\ding{52}\ding{52} & \ding{52} & \ding{52}\ding{52}\ding{52}& \ding{52}\ding{52} & \ding{52}\ding{52}\ding{52} & \ding{52}  & \ding{55} & \ding{55}  \\
MAXI J1659-352 & \ding{55} & \ding{55} & \ding{55} & \ding{55}  & \ding{55} & \ding{55} & \ding{55} & \ding{55}  \\
\hline
Matched tracks & 13/15 & 9/15 & 13/15 & 10/15 & 13/15 & 11/15 & 8/15 & 7/15 \\
\enddata
\end{deluxetable*}

\subsection{Formatioin of BH UCXBs}

Ultracompact X-ray binaries (UCXBs) are a subclass of LMXBs characterized by ultra-short orbital periods (less than about 1 hour) and hydrogen-deficient donor stars \citep{Nelson1986,Nelemans2010}. To date, approximately 45 UCXBs and candidates have been identified, but no BH UCXBs have been observed so far \citep{ArmasPadilla2023}. 

Figure 2 indicates that, with all of the four MB laws it is possible to form BH UCXBs with orbital periods less than 1.5 hours. Our calculations suggest that BH UCXBs can form in two channels. The corresponding examples are shown in Figure 4 with an initial BH mass of 8\,$M_{\odot}$. The first category of BH UCXBs contain a He WD companion. The initial orbital period is above the bifurcation period so the donor can evolve to be a He WD and detache from its RL at the orbital period around 10 hours. Then the binary orbit continues to shrink under the influence of GR. This eventually causes the He WD to fill the RL within the Milky Way's lifetime, enabling mass transfer once more at a rate as high as $\sim 10^{-6}\,M_{\odot}$\,yr$^{-1}$, significantly  exceeding the BH's Eddington accretion rate. Thus, this type of UCXBs may be observed as ultraluminous X-ray sources. The solid line represents such an evolution with initial donor mass of 1.4\,$M_{\odot}$, orbital period of 0.9 day, and the VI19 law. 
The second category contain either a very low-mass MS star or a brown dwarf companion with mass less than 0.08\,$M_{\odot}$. The dotted line denotes the evolution with initial donor mass of 1.0\,$M_{\odot}$, orbital period of 0.5 day (which is below the bifurcation period), and the RVJ83 law. Mass transfer initiates when the donor is on main sequence, and the donor retains its original hydrogen (H) and helium (He) abundances during the evolution. The minimum orbital period is around 1 hour and the mass transfer rate is less than $10^{-9} M_{\odot}$ yr$^{-1}$.

\begin{figure}
    \centering
    \includegraphics[width=18cm]{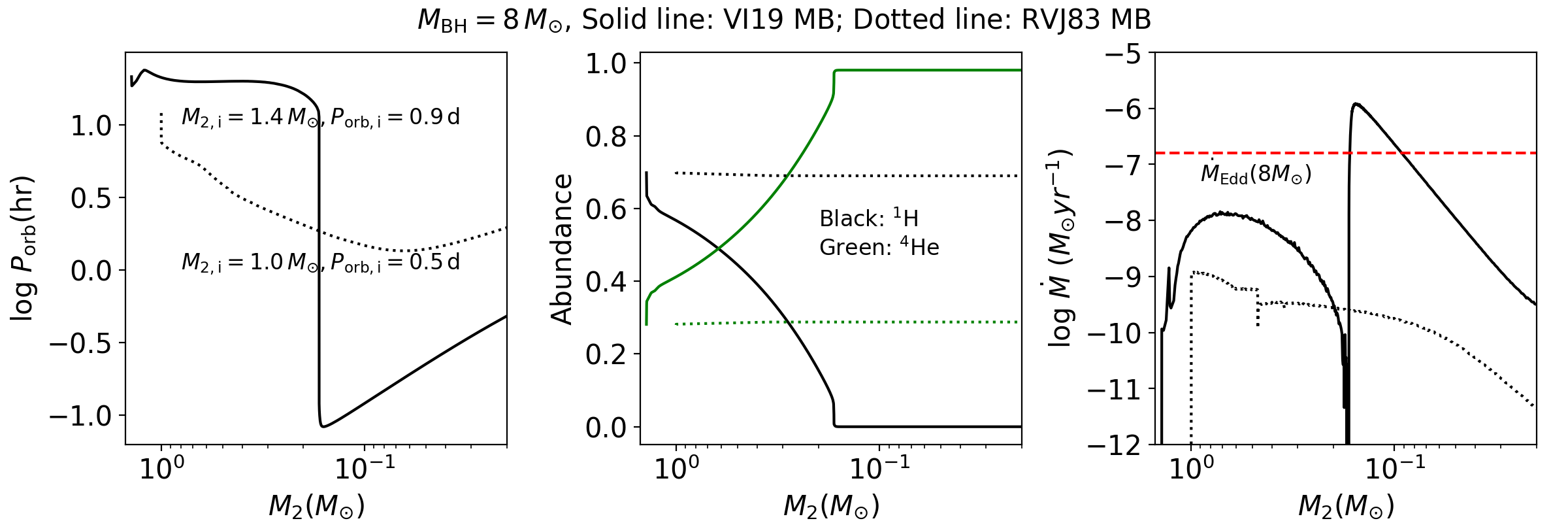}
    \caption{Examples of the formation of BH UCXBs with orbital periods less than 90 minutes. The left, middle, and right panels show the evolution of the orbital period, H and He abundances in the companion star, and mass transfer rate versus the companion mass, respectively. The initial BH mass is $8\,M_{\odot}$. The solid line represents the evolution with an initial companion mass of $1.4\,M_{\odot}$ and an initial orbital period of 0.9 day using the VI19 law. The dotted line represents the evolution with an initial companion mass of $1.0\,M_{\odot}$ and an initial orbital period of 0.5 day using the RVJ83 model. The black and green colors in the middle panel demonstrate the hydrogen and helium abundances of the companion, respectively. The red dashed line in the right panel represents the Eddington accretion rate.}

\end{figure}

\section{Binary Population Synthesis}
In our binary evolution calculations in last section we have adopted uniform distributions in the initial parameters of BH I/LMXBs. However,
the actual number of BH I/LMXBs evolved with specific initial parameter combination depends on the binary formation rate within the parameter space, which is determined by the initial conditions of the primordial binaries, and the physical processes like mass transfer, common envelope (CE) evolution and supernova (SN) explosion.
In this section, we follow the formation and evolution of BH LMXBs by combining the BPS with detailed evolutionary calculations.

\subsection{BPS model and initial parameters}

We use the BSE code originally developed by \citet{Hurley2002} to follow evolution of the progenitor binaries of BH LMXBs in the Milky Way. We adopt a star formation rate of 3 $M_{\odot}$ yr$^{-1}$ and a constant metallicity of $Z=0.02$ throughout the 12 Gyr lifespan of the Milky Way galaxy. The primordial binaries consist of a primary of mass $M_1$ and a secondary of mass $M_2$ in a circular orbit with a separation $a$. We assume that the primary mass distribution follows the initial mass function of \citet{Kroupa1993}, with a mass range of $19-60M_{\odot}$. According to \citet{Kobulnicky2007}, we assume that the mass ratio ($q=M_2/M_1$) is uniformly distributed between 0 and 1. We also assume that the initial orbital separation $a$ is uniformly distributed in the logarithmic range of $3-10^4 R_{\odot}$ \citep{Abt1983}.

We adopt the fitting formula of \citet{Nieuwenhuijzen1990} to model stellar wind mass loss. For hot OB stars with effective temperature $T_{\rm eff}>11000$ K and stripped helium stars, we replace it with the simulated relations by \citet{Vink2001} and \citet{Vink2017}, respectively. We assume that half of the transferred mass is accreted by the secondary, and the excess material is assumed to be ejected out of the binary in the form of isotropic winds, taking away the accretor's specific angular momentum \citep{Deng2024}. We adopt the critical mass ratios $q_{\rm cr}$ calculated by \citet[][their Figure~1]{Shao2014} to determine the stability of the mass transfer. If the mass transfer is dynamically unstable, a CE phase follows. The standard energy conservation equation \citep{Webbink1984} is employed to deal with CE evolution. We set the CE evolution energy efficiency $\alpha=1.0$, and use the calculated binding energy parameters $\lambda$ of the primary's envelope in \citet{Xu2010} and \citet{Wang2016RAA}.

For the formation of BHs, we take into account four SN models to treat the remnant masses and natal kicks: 
\begin{enumerate}
  \item SN A - the rapid explosion model \citep{Fryer2012},
  \item SN B - the delayed explosion model \citep{Fryer2012},
  \item SN C - the stochastic explosion model \citep{Mandel2020},
  \item SN D - the failed SN explosion model \citep{O'Connor2011,Wang2016MN}.
\end{enumerate}
The mass of the compact objects in SN A and B models is determined by the mass of the CO core at the time of the explosion. Note that SN A model cannot form compact objects with mass $\sim 2-5\,M_{\odot}$. The Failed SN model is based on the hypothesis that the formation of BH is determined by the compactness of the stellar core at the time of the collapse: low-compactness cores produce NSs through SNe, while high-compactness cores experience failed SNe and produce BHs \citep{O'Connor2011}. The BH mass is taken to be the He core mass before SNe, $M_{\rm BH}=M_{\rm He,preSN}$ \citep[e.g.,][]{Kochanek2014,Kochanek2015,Smith2011, Shiode2014,Smith2014,Sukhbold2014,Clausen2015}. 

For the BHs formed by the rapid, delay, and failed SN models, we assume that the natal kick velocity is inversely proportional to the BH mass: $V_{\rm kick}({\rm BH})=(3\,M_{\odot}/M_{\rm BH})V_{\rm kick}({\rm NS})$, where $V_{\rm kick}({\rm NS})$ is the kick velocity of NSs, which follows a Maxwell distribution with $\sigma=265$\,km\,s$^{-1}$. For the stochastic SN model, the mass and kick velocity satisfy a specific probability distribution based on the mass of its CO core \citep{Mandel2020}. Similar to \citet{Fryer2012}, we simply assume that the gravitational mass of the BH is 90\% of its baryonic mass. For all the above SN models, we assume that a compact object with a mass greater than 3$M_{\odot}$ is a BH, and do not consider BHs formed from accretion-induced collapse of NSs.

\subsection{Incipient BH+MS binaries}

Figure 5 shows the obtained distribution of the orbital period and companion mass of incipient BH+MS binaries in different SN models. In each model, we generate several million incipient BH+MS binaries based on the BPS calculations. We then select the systems with orbital periods shorter than 1000 days and companion masses less than $10\,M_{\odot}$. In SN A model, the BH progenitors are initially more massive than $\sim 25\,M_{\odot}$. Consequently, the companion stars less massive than $\sim 3\,M_{\odot}$ are unable to expel the star's envelope during CE evolution \citep[see also][]{Podsiadlowski2003,Wang2016MN,Shao2019}, resulting in merger of both stars. In comparison, the BH progenitor masses in other SN models can be considerably lower, thus the companion stars with mass $\lesssim 1.5\,M_{\odot}$ can survive CE evolution. The birth rates of the incipient BH binaries are $8.4\times 10^{-7}$, $1.2\times 10^{-6}$, $8.0\times 10^{-6}$, and $7.3\times 10^{-6}$\,yr$^{-1}$ in SN A-D models, respectively.

\begin{figure}
    \centering
    \includegraphics[width=10cm]{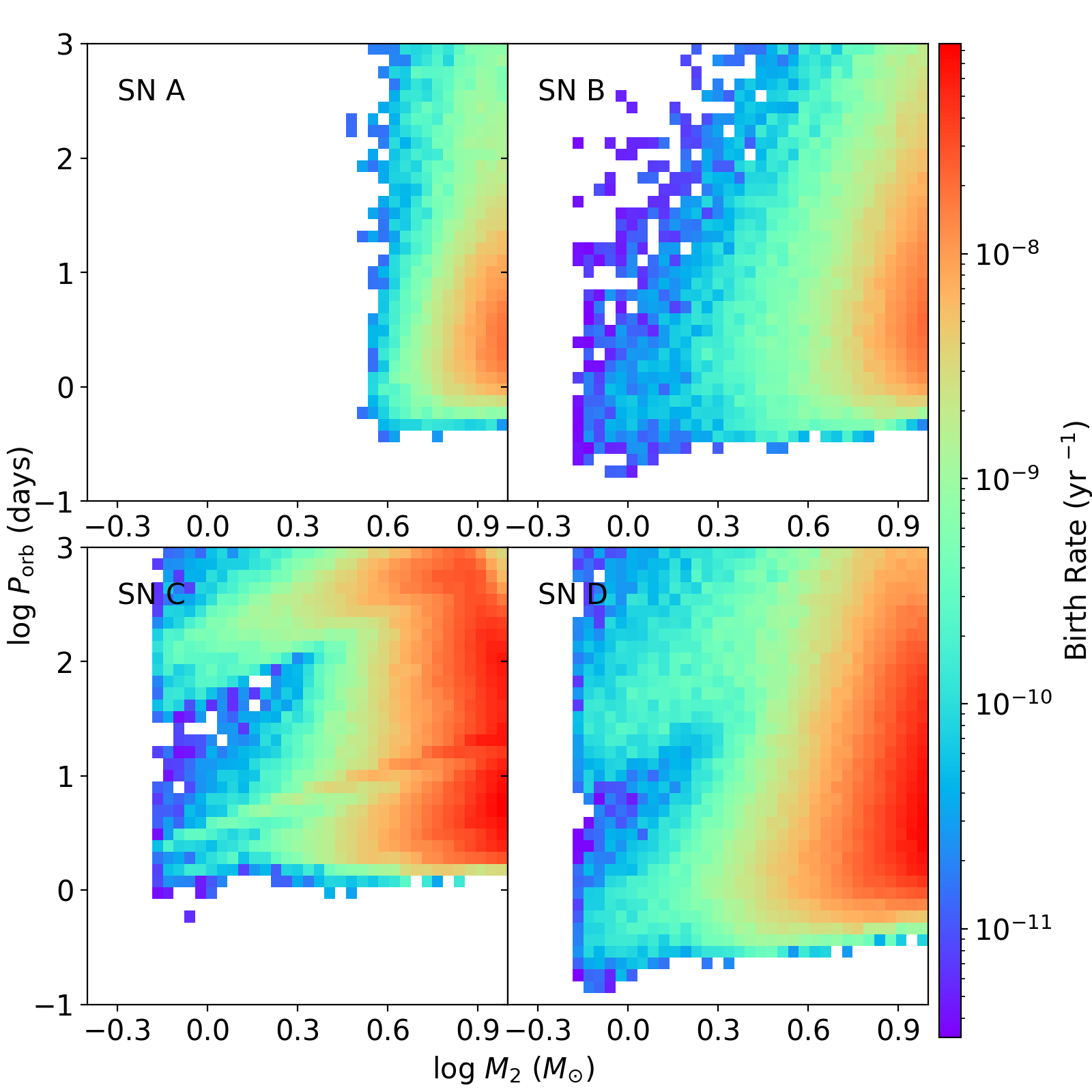}
    \caption{The distribution of the companion mass and orbital period of incipient BH+MS binaries. We use different colors to represent the magnitude of the birth rate. 
}

\end{figure}

\subsection{The orbital period and accretion rate distribution of  Galactic BH LMXBs}

We combine the birth rate of incipient BH+MS binaries from BPS with the detailed evolution of BH I/LMXBs calculated with MESA to obtain the population properties of Galactic BH LMXBs. Specifically, we calculate the birth rate in each ($\Delta M_{\rm BH}$, $\Delta M_{\rm c}$, $\Delta P_{\rm orb}$) grid, and multiply it by the evolutionary time to obtain the number and distribution of all BH LMXBs, from which we select systems with the mass transfer rate $\dot{M}_{\rm tr}>10^{-13} M_{\odot}$ yr $^{-1}$.

Figure 6 shows the distribution of BH LMXBs in the orbital period and accretion rate plane. We also plot the observational data for comparison. BH LMXBs with orbital periods between 0.1 and 0.6 day occupy more than half of the total population, and their mean accretion rates are all less than $10^{-9}\,M_{\odot}$\,yr$^{-1}$. Since the evolution of LMXBs with orbital periods in this range is mainly driven by MB, comparison between theory and observation can provide useful probe to both the SN models and the MB laws. 

The top panel of Figure 6 reveals that no LMXBs can form with orbital periods shorter than $\sim 0.6$ day in SN A model, inconsistent with observations. So this model can be ruled out. The other three models (SN B-D) can produce systems with orbital periods shorter than 0.6 day, and the expected numbers of BH LMXBs increase in the order of SN B, SN D, and SN C (from $\sim 40-50$ in SN B model to $\sim 400$ in SN C model, see Table 3). 
By analysing the distances and spatial distribution of the observed BH LMXBs, \citet{Corral2016} estimated a total population of $\sim 1280(\frac{{\rm ORP}}{100\,{\rm yr}})$, where ORP is the mean outburst recurrence period. This means that SN C and D models are more ready to account for the total number of the BH LMXBs in the Milky Way galaxy.

To demonstrate the impact of different MB laws on the accretion rates, we compare in Figure 7 the calculated and observed accretion rate distributions for BH LMXBs with the orbital periods between 0.1 and 0.6 day in SN B, C and D models. We see that the accretion rates calculated with the RVJ83 law seem to best match the observations, mainly distributed between $\sim 10^{-10}$ M$_{\odot}$ yr$^{-1}$ and $\sim 10^{-9}$ M$_{\odot}$ yr$^{-1}$. The peak accretion rate ($\sim$ a few $10^{-10}$ M$_{\odot}$ yr$^{-1}$) with the RM12 law is also close to the observation, but the number of the systems with the accretion rates above $10^{-9.5}$ M$_{\odot}$ yr$^{-1}$ is relatively small. In comparison,  the VI19 and $\tau-$boosted laws predict a wider distribution of the accretion rate and more large-accretion rate ($\geq 10^{-9}$ M$_{\odot}$ yr$^{-1}$) LMXBs than observations.

\begin{figure}[htb]
    \centering
    \includegraphics[width=18cm]{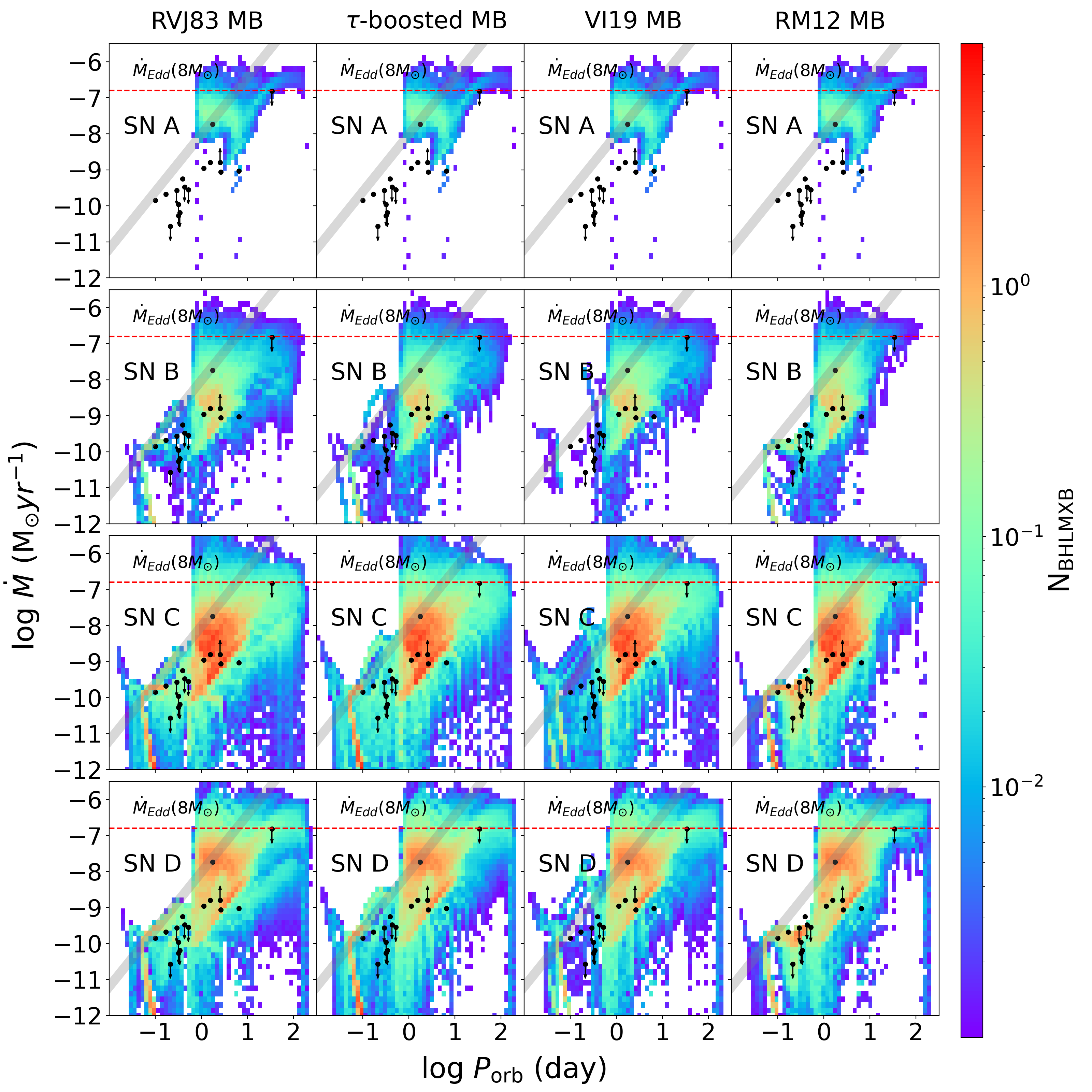}
    \caption{Comparison of the calculated and observed distributions of the orbital period and accretion rate of BH LMXBs in different SN and MB models. The red dashed line represents the Eddington accretion rate of a $8\,M_{\odot}$ BH, and the gray thick line represents the critical accretion rate for disk instability. The black dots depict the distribution of the observed BH LMXBs.
}
\end{figure}

\begin{figure}[htb]
    \centering
    \includegraphics[width=18cm]{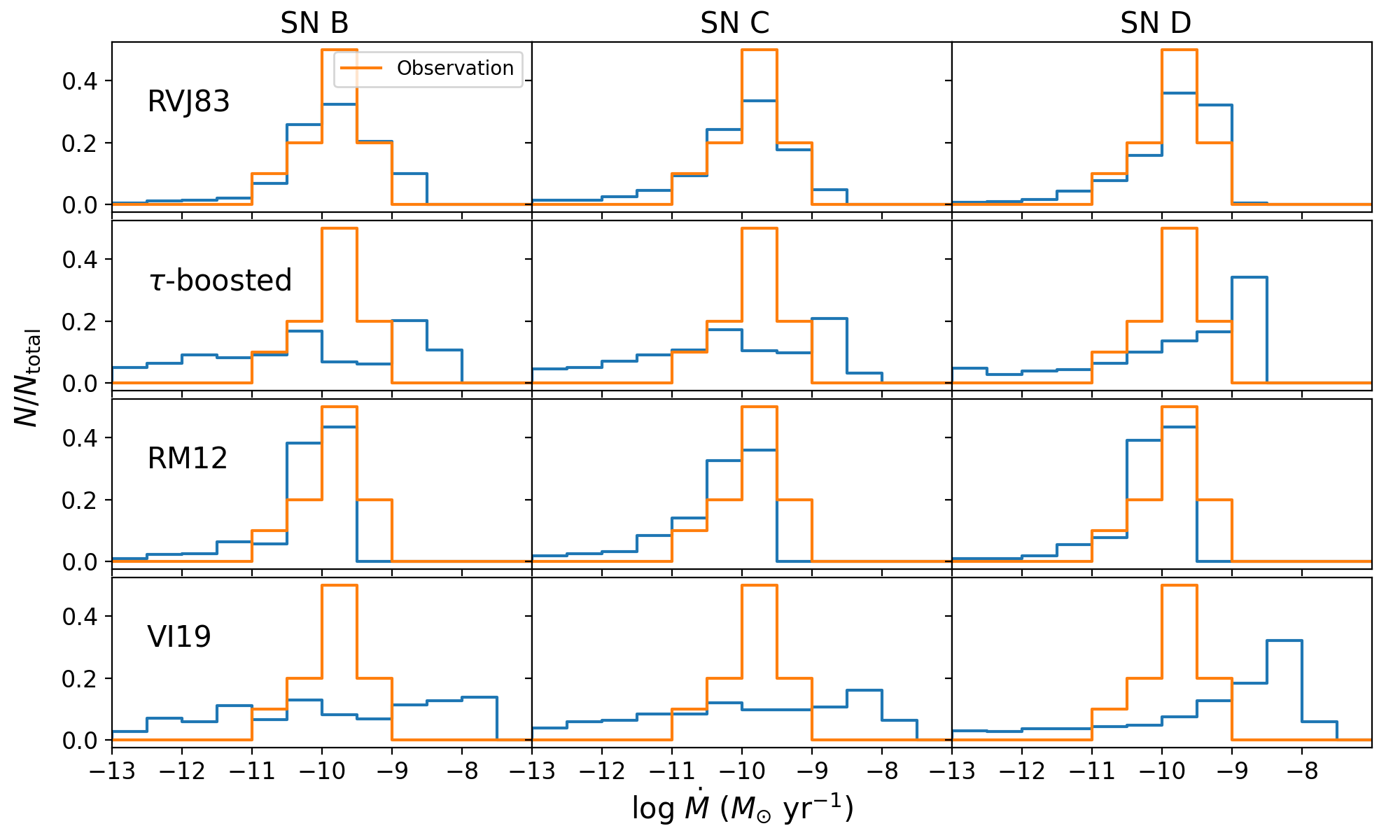}
    \caption{Accretion rate distribution of BHLMXBs with calculated and observed orbital periods between 0.1 and 0.6 days. The blue, orange, green, red, and purple lines represent the results calculated using the Skumanich model, $\tau-$boosted model, RM12 model, VI19 model, and observed values, respectively..
}
\end{figure}

We then examine the orbital period distribution of BH LMXBs.
We classify them into three categories: UCXBs with $P_{\rm orb}<90$ minutes, compact LMXBs with 90 minutes $<P_{\rm orb}<1$ day, and wide LMXBs with $P_{\rm orb}>1$ day. We choose 1 day as the dividing period between compact and wide binaries because it is close to the so-called bifurcation period of LMXBs \citep{Pylyser1988,Pylyser1989,Podsiadlowski2002}. LMXBs with initial orbital periods longer than the bifurcation period will expand due to mass transfer, while binaries with shorter initial orbital periods will shrink due to MB. Table 3 lists the percentages and total numbers of the three classes of LMXBs in different SN and MB models. The last row presents the observed data. The predicted number of BH LMXBs is not sensitively dependent on the MB laws, except slightly relatively small number with the VI19 law. 

\citet{Belczynski2004} and \citet{vanHaaften2013} investigated the formation of UCXBs. All UCXBs in their simulations have a NS accretor, without any UCXBs with a BH formed directly in the collapse of a massive star. We reach the same conclusion in SN A model, which predicts that binaries with a very massive primary will merge during the unstable mass transfer stage. However, in SN B-D models, there do exist BH UCXBs regardliess of the MB laws, but their percentage in the BH LMXB population is negligible ($\lesssim 2.5\%$). 
This is basically compatible with the fact that no BH UCXBs have been discovered yet. In comparison, there are around 45 NS UCXBs and candidates identified \citep{ArmasPadilla2023}. This sharp contrast may originate from their formation channels. Most BH UCXBs evolve from BH LMXBs, while NS UCXBs evolve from both NS LMXBs and IMXBs. In particular, the latter can efficiently produce NS+WD binaries through CE evolution, which further evolve to NS UCXBs. The birth rate of NS IMXBs are significantly higher than that of BH LMXBs \citep{Pfahl2003}, leading to a much larger population of NS UCXBs compared to BH UCXBs.

In SN B-D models, BH LMXBs are predicted to be dominated by wide rather compact binaries. This seems to be in conflict with observations, which show that about $65\%$ BH LMXBs are compact binaries. Since most BH LMXBs are expected to be transient (see Figure 6), it could be partially caused by the selection effect, i.e., wider LMXBs spend longer time in quiescence \citep{Lasota2008}, and are more difficult to detect.

\begin{deluxetable*}{cccccc}
\tablenum{3}
\tablecaption{The number of BHL/IMXB systems formed by different models}
\tabletypesize{\scriptsize}
\tablehead{
 SN & MB model & $R_{\rm BHUCXB}$ & $R_{\rm BHLMXB}$ & $R_{\rm BHI/LMXB}$ & $N_{\rm BHL/IMXRB}$  \\
 & & ($P_{\rm orb}<90$\,min) & ($90\,{\rm min}<P_{\rm orb}<1$\,d) & ($P_{\rm orb}>1$\,d) 
}
\startdata
\multirow{4}{*}{A} & Skumanich &       0.0 & 7.4\% & 92.6\% & 6.6 \\
 & $\tau-$boosted &  0.0  & 7.4\% & 92.6\% & 6.6 \\
 & VI19 &            0.0  & 7.4\% & 92.6\% & 6.6 \\
 & RM12 &            0.0  & 7.4\% & 92.6\% & 6.6 \\
\hline
\multirow{4}{*}{B} & Skumanich &       1.6\% & 25.7\% & 72.7\% & 52.6 \\
 & $\tau-$boosted &  1.6\% & 27.6\% & 70.8\% & 52.7 \\
 & VI19 &            0.5\% & 14.6\% & 84.9\% & 43.8  \\ 
 & RM12 &            1.2\% & 26.7\% & 72.1\% & 50.7 \\
\hline
\multirow{4}{*}{C} & Skumanich &       1.1\% & 26.1\% & 72.8\% & 395 \\ 
 & $\tau-$boosted &  1.7\% & 28.8\% & 69.5\% & 399.8 \\
 & VI19 &            1.0\% & 17.5\% & 81.5\% & 334.7  \\ 
 & RM12 &            0.9\% & 29.0\% & 70.1\% & 388.5\\ 
\hline
\multirow{4}{*}{D} & Skumanich &       1.5\% & 26.6\% & 71.9\% & 225.8 \\ 
 & $\tau-$boosted &  2.3\% & 31.4\% & 66.3\% & 238.8 \\
 & VI19 &            2.3\% & 16.6\% & 81.1\% & 193.9  \\ 
 & RM12 &            1.3\% & 28.9\% & 69.8\% & 223.4 \\ 
\hline
\multicolumn{2}{c}{Observation} & 0.0 & 65.2\% & 34.8\% & 23 \\
\enddata

\end{deluxetable*}

\section{Concluding remarks}

In this work, we combine BPS and detailed stellar evolution to follow the formation and evolution of BH LMXBs and examine the viability of different MB laws. We select 15 BH LMXBs with measured orbital periods, BH masses, companion masses, temperatures, and mean accretion rates, and compare them with the simulated results. We find that 
the RVJ83 and RM12 MB laws seem to be in better agreement with the observed properties of BH LMXBs compared with other two MB laws, regardless of the SN models.




\citet{Deng2021} examined the the applicability of the five MB laws for NS LMXBs, and found that the more efficient ones, i.e., the VI19 and $\tau$-boosted laws are more successful in explaining the persistent, luminous NS LMXBs, the orbital period distribution of binary pulsars, and the formation of NS UCXBs \citep[see also][]{Chen2021,Echeveste2024}. \citet{Gossage2023} suggested 
the \citet{Garraffo2018} prescription, which incorporates  magnetic saturation like the RM12 law, is capable of 
modeling the spin period and mass data from open
clusters and NS LMXBs, but they did not make comparison of the accrete rate data and calculate the population densities of the LMXB models. By constraining the orbital period distribution of low-mass detached MS eclipsing binaries, \citet{El-Badry2022}
advocated using saturated MB laws in binary evolution calculations in tidally locked binaries like cataclysmic variables (CVs). They also pointed out that such models face difficulties in explaining the period gap and the mass transfer rates in CVs compared with the RVJ83 model.

Since most CVs, BH and NS LMXBs share similar properties of the orbits and companion stars, as well as the mass transfer process, it is enigmatic that their evolutions obey different MBs laws. As mentioned earlier, it is possible that all of the adopted MB laws lack certain crucial components and some `mixture' of them might be required. We anticipate the development of a more universal MB law in the future that can comprehensively explain the observational data of both single low-mass stars and mass-transferring binaries.

Here we propose several additional possibilities to explain why similar binaries prefer different MB laws. Firstly, the MB law may be actually the same, but there are additional source(s) of angular momentum loss associated with mass loss. Depending on the amount of the lost mass and of the carried specific angular momentum, the efficiency of angular momentum loss could vary by a large factor in different types of mass-transferring binaries \citep{Soberman1997}.
Secondly, the discrepancy might originate from the different extent of X-ray irradiation on the companion star. X-ray irradiation is relatively weak in CVs because of the low X-ray luminosities of accreting white dwarfs, but may significantly change the inner structure of the companion star and/or drive ablation of the companion star's outer layers in NS LMXBs \citep[e.g.,][]{Podsiadlowski1991,Buning2004,Benvenuto2015,Tailo2018}. While nearly half of the observed NS LMXBs are persistent sources, most BH LMXBs appear as X-ray transients, occasionally undergoing a transition from quiescent to outburst states. Compared to NSs, BHs do not have a hard surface, so the X-ray spectra during outbursts are relatively soft \citep{Remillard2006}.
Thus, the inner structure and atmosphere of the companion star are likely less affected by X-ray irradiation in BH LMXBs than in NS LMXBs. Clearly, more observational and theoretical works are needed to reconcile the discrepant MB laws inferred from both single stars and different types of interacting binaries.

\begin{acknowledgements}
We are grateful to the referee for helpful comments. This work was supported by the National Key Research and Development Program of China (2021YFA0718500), the Natural Science Foundation of Chian under grant No. 12041301 and 12121003.
\end{acknowledgements}

\bibliographystyle{aasjournal}
\bibliography{BHLMXB}{}


\end{document}